\documentclass[%
superscriptaddress,
preprintnumbers,
nofootinbib,
 amsmath,amssymb,
 twocolumn,
]{revtex4-1}
\usepackage{aas_macros,braket}
\usepackage{bm}
\usepackage{dcolumn}
\usepackage{graphicx}
\usepackage{natbib}

\begin{document}

%
\title{Measuring primordial anisotropic correlators with CMB spectral distortions}
\author{Maresuke Shiraishi}
\affiliation{%
Kavli Institute for the Physics and Mathematics of the Universe (Kavli IPMU, WPI), UTIAS, The University of Tokyo, Chiba, 277-8583, Japan
}
\affiliation{%
Dipartimento di Fisica e Astronomia ``G. Galilei'', Universit\`a degli Studi di Padova, via Marzolo 8, I-35131, Padova, Italy
}
\affiliation{%
INFN, Sezione di Padova, via Marzolo 8, I-35131, Padova, Italy
}
\author{Michele Liguori}
\affiliation{%
Dipartimento di Fisica e Astronomia ``G. Galilei'', Universit\`a degli Studi di Padova, via Marzolo 8, I-35131, Padova, Italy
}
\affiliation{%
INFN, Sezione di Padova, via Marzolo 8, I-35131, Padova, Italy
}
\author{Nicola Bartolo}
\affiliation{%
Dipartimento di Fisica e Astronomia ``G. Galilei'', Universit\`a degli Studi di Padova, via Marzolo 8, I-35131, Padova, Italy
}
\affiliation{%
INFN, Sezione di Padova, via Marzolo 8, I-35131, Padova, Italy
}
\author{Sabino Matarrese}
\affiliation{%
Dipartimento di Fisica e Astronomia ``G. Galilei'', Universit\`a degli Studi di Padova, via Marzolo 8, I-35131, Padova, Italy
}
\affiliation{%
INFN, Sezione di Padova, via Marzolo 8, I-35131, Padova, Italy
}
\affiliation{%
Gran Sasso Science Institute, INFN, viale F. Crispi 7, I-67100, L'Aquila, Italy
}

\date{\today} \preprint{IPMU15-0090}
\begin{abstract}
  We show that inflationary models with broken rotational invariance generate testable off-diagonal signatures in the correlation between the $\mu$-type distortion and temperature fluctuations of the cosmic microwave background. More precisely, scenarios with a quadrupolar bispectrum asymmetry, usually generated by fluctuations of primordial vector fields, produce a nonvanishing $\mu$-$T$ correlation when $|\ell_1-\ell_2|=2$. Since spectral distortions are sensitive to primordial fluctuations up to very small scales, a cosmic variance limited spectral distortion experiment can detect such effects with a high signal-to-noise ratio. 
\end{abstract}
\maketitle

\section{Introduction}

Deviations of the cosmic microwave background (CMB) frequency spectrum from a perfect blackbody distribution, generally referred to as 
``CMB spectral distortions,'' can be an important source of information about the thermal history of the early Universe.
Spectral distortions are in effect generated by any process that could inject energy in 
the baryon-photon plasma, starting from very early times, and can thus be used to study and constrain such processes.
At present we have no detection of CMB spectral distortions, with the best constraint dating back to the COBE/FIRAS limit, 
${\Delta I_{\nu}/I_{\nu}} \leq 10^{-4}$ \cite{Fixsen:1996nj, Mather:1993ij}.~\footnote{More specifically, we have a detection, coming from {\it Planck}, ACT and SPT data, of the fluctuating part of the $y$-type spectral distortion, generated by clusters and by the intergalactic medium (see, e.g., Ref.~\cite{Khatri:2015jxa}), but we have neither a detection 
of the invariant part of $y$ distortion, nor of $\mu$ distortions, generated at early times, via dissipation of acoustic waves, or during reionization.} 
However, current technological advancements and planned space missions, such as PIXIE \cite{Kogut:2011xw}, will produce dramatic improvements in the near 
future. For these reasons, spectral distortions have been receiving increasing attention in recent years, and they are 
the starting point for a growing number of new and original applications.

In this context, it was recently pointed out in  
Ref.~\cite{Pajer:2012vz} that CMB spectral distortions from dissipation of acoustic waves in the primordial plasma can be used to 
 test the statistical properties of primordial cosmological perturbations. More specifically, the authors of Ref.~\cite{Pajer:2012vz} 
show that the correlation between the so-called $\mu$ distortion parameter and the CMB temperature fluctuations probes the  
three-point function (bispectrum) of primordial curvature perturbations in the squeezed limit (i.e. for triangle configurations 
in which one wave number is much smaller than the other two, thus producing a coupling between small and large scales). 
High-precision measurements of the 
$\mu$-$T$ correlation can then be used to estimate the bispectrum amplitude parameter $f_{\rm NL}$ for models 
predicting non-Gaussianity (NG) signals with a significant contribution from squeezed triangles. 
Similarly, the $\mu$-$\mu$ autocorrelation probes the trispectrum of primordial fluctuations and can be used to infer the $\tau_{\rm NL}$ parameter~\cite{Pajer:2012vz}. The main interest of this approach to testing primordial NG, in contrast to other methods such as measurements of the CMB temperature bispectrum, is that spectral distortions allow one to reach much smaller 
scales than what was achievable with temperature anisotropies, where $\ell_{\rm max}$ is limited by Silk damping. This in turn 
 produces a significant increase of sensitivity for a cosmic-variance-dominated experiment. Different groups have so far 
studied the possibility to use $\mu$ distortions in order to constrain primordial NG of the local type \cite{Pajer:2012vz,Ganc:2012ae}, 
eventually including a running of the $f_{\rm NL}$ parameter \cite{Biagetti:2013sr, Emami:2015xqa}, NG generated by models with excited initial states \cite{Ganc:2012ae}, isocurvature modes \cite{Ota:2014iva} or primordial magnetic fields \cite{Miyamoto:2013oua, Kunze:2013uja, Ganc:2014wia}.

 In this paper, we will instead consider primordial scenarios characterized by the breaking of rotational 
invariance. We consider both Gaussian and non-Gaussian primordial perturbation fields, introducing, respectively, a spatial modulation in the power spectrum and a preferred direction in the bispectrum.  Anisotropic bispectra can arise in inflationary models with a U(1) gauge field (e.g., Refs.~\cite{Bartolo:2011ee,Bartolo:2012sd,Bartolo:2015dga}), while Gaussian primordial perturbations 
characterized by power spectra with a dipolar modulation have been considered in light of the observed hemispherical power 
asymmetry in CMB data \cite{Hoftuft:2009rq, Ade:2013nlj, Akrami:2014eta, Ade:2015sjc, Aiola:2015rqa}. The vast majority of inflationary models with breaking of rotational invariance predict primordial bispectra which peak in the squeezed limit. It is thus reasonable to expect these bispectra to generate non-negligible 
$\mu$-$T$ correlations. This is indeed what we will find. Moreover, the anisotropic quadrupolar nature of the primordial signal produces off-diagonal couplings, sourcing 
terms with $|\ell_1-\ell_2|=2$.
Besides anisotropic inflationary scenarios, which are the main 
 focus of this paper, we also consider phenomenological 
models of dipolar modulation, which were originally motivated by the observed power asymmetry in CMB data.  
Since a trispectrum signal of the $\tau_{\rm NL}$ type gives rise to large-scale modulation of small-scale power, and this leaves a signature in the $\mu$-$\mu$ correlation function, 
we can expect also these dipolar modulation models to generate $\mu$-$\mu$ correlations. We will show that, if a modulation shows up at very small scales, a signal of this type is actually generated. 
Also in this case, the dipolar anisotropy sources off-diagonal $\mu$-$\mu$ couplings, this time with $|\ell_1-\ell_2|=1$.
The observed CMB asymmetry is, however, only on large scales, $\ell \lesssim 60$ \cite{Hoftuft:2009rq, Ade:2013nlj, Akrami:2014eta, Ade:2015sjc, Aiola:2015rqa}. In this case, we expect the signal to vanish, since spectral distortions test only very small scales. Our results will confirm this intuitive expectation.

This paper is organized as follows: In Sec.~\ref{sec:CMBmu} we briefly summarize the physical mechanism which generates $\mu$-type spectral 
distortions; in Secs.~\ref{sec:anisNG} and \ref{sec:dipolarpower} we derive the $\mu$-$T$ and $\mu$-$\mu$ correlation functions, respectively, for 
models with anisotropic NG and dipolar power asymmetry. Some Fisher matrix forecasts for ideal cosmic-variance-dominated experiments and more realistic experiments are shown in Sec.~\ref{sec:Fisher}. Finally, we summarize our main results and draw our conclusions in Sec.~\ref{sec:conclusions}.

\section{CMB $\mu$ distortion anisotropy}\label{sec:CMBmu}

Heating due to diffusion damping of acoustic waves can induce distortions in the CMB blackbody spectrum. At very early times, when double Compton scattering is efficient ($z \gtrsim z_i \equiv 2 \times 10^6$), any generated distortion is immediately erased due to quick thermalization, and the blackbody spectrum is maintained. On the other hand, for $z_f \lesssim z \lesssim z_i$ with $z_f \equiv 5 \times 10^4$, double Compton scattering has become inefficient, and Compton scattering mainly thermalizes the system, without changing the number of photons. Hence, due to energy injection via acoustic wave dissipation, a blackbody spectrum is altered to a Bose-Einstein distribution with nonzero chemical potential: \cite{Sunyaev:1970er, 1975SvA....18..413I, 1982A&A...107...39D, 1991A&A...246...49B, Hu:1994bz, Chluba:2011hw, Khatri:2011aj, Chluba:2012gq, Khatri:2012tv, Khatri:2012tw}
\begin{eqnarray}
\mu \approx \frac{1.4}{4} 
\left[ \Braket{\delta_\gamma^2(z)}_p\right]_{z_f}^{z_i} ~,
\end{eqnarray}
where $\braket{}_p$ denotes a quantity averaged over an oscillation period, and $\left[ F(z) \right]_{z_f}^{z_i} \equiv F(z_i) - F(z_f)$. The density fluctuation affecting $\mu$ is expressed as (see e.g., Refs.~\cite{1968ApJ...151..459S, 1970ApJ...162..815P, 1983MNRAS.202.1169K,Weinberg:2008zzc})
\begin{eqnarray}
\Braket{\delta_\gamma^2({\bf x})}_p 
&\simeq& 1.45 \left[ \prod_{n=1}^2 \int \frac{d^3 {\bf k}_n}{(2\pi)^3} 
\zeta_{{\bf k}_n} \right] 
\nonumber \\ 
&& \Braket{\Delta_\gamma(k_1) \Delta_\gamma(k_2) }_p
e^{i ({\bf k}_1 + {\bf k}_2) \cdot {\bf x}} ~,
\end{eqnarray}
where $\zeta$ denotes primordial curvature perturbation and the photon transfer function is approximately given by $\Delta_\gamma(k) \simeq 3 \cos (k r) \exp[ -k^2 / k_D^2(z) ]$ with $r(t) \simeq 2t / (a \sqrt{3})$. The diffusion damping scales at $z_i$ and $z_f$ are, respectively, $k_D(z_i) \approx 12000 ~{\rm Mpc^{-1}}$ and $k_D(z_f) \approx 46 ~{\rm Mpc^{-1}}$ \cite{1968ApJ...151..459S, 1970ApJ...162..815P, 1983MNRAS.202.1169K, Weinberg:2008zzc}.

The above equations yield an expression for the $\mu$ distortion \cite{Pajer:2012vz, Ganc:2012ae}:
\begin{eqnarray}
\mu({\bf x}) &\simeq& 
\left[ \prod_{n=1}^2 
\int \frac{d^3 {\bf k}_n}{(2\pi)^3} 
\zeta_{{\bf k}_n} 
\right] 
\int d^3 {\bf k}_3 
 \delta^{(3)}\left(\sum_{n=1}^3 {\bf k}_n \right) \nonumber \\ 
&& f(k_1, k_2, k_3)
e^{-i {\bf k}_3 \cdot {\bf x}}  \label{eq:mu} ~, \\
f(k_1, k_2, k_3) &\equiv& \frac{9}{4} W\left(\frac{k_3}{k_s}\right) 
\left[ e^{-(k_1^2 + k_2^2)/k_D^2(z)} \right]_{z_f}^{z_i} ~,
\end{eqnarray}
where $W(x) \equiv 3 j_1(x) / x$ is a top-hat filter function, limiting the scales where the acoustic waves are averaged to give heat as $k_3 / k_s \lesssim 1$, and we have used an estimate: $\Braket{\cos(k_1 r) \cos(k_2 r)}_p \simeq 1/2$. In the following analysis, we take $k_s$ to be $k_D(z_f)$, in order to estimate a lower bound on the $\mu$ distortion. Then, the transfer function $f(k_1, k_2, k_3)$ filters the squeezed-limit signals, satisfying $k_1, k_2 > k_D(z_f) > k_3$. As seen in Eq.~\eqref{eq:mu}, the $\mu$ distortion has quadratic dependence on the primordial curvature perturbation (while usual CMB anisotropy is simply proportional to the curvature perturbation), and hence the correlation between the $\mu$ distortion and the CMB temperature anisotropy or the power spectrum of the $\mu$ distortion is generated if the bispectrum or trispectrum of curvature perturbations is nonzero \cite{Pajer:2012vz, Ganc:2012ae}. In what follows, we analyze these correlators, including broken rotational invariance.

\section{$\mu$-$T$ correlation due to primordial anisotropic non-Gaussianity}\label{sec:anisNG}

We here investigate potential imprints on spectral distortions of primordial models which generate a curvature bispectrum with a quadrupolar asymmetry. 
This can be expressed as 
$\Braket{\zeta_{{\bf k}_1} \zeta_{{\bf k}_2} \zeta_{{\bf k}_3}} 
= (2\pi)^3 \delta^{(3)}\left({\bf k}_1 + {\bf k}_2 + {\bf k}_3 \right) B_{{\bf k}_1 {\bf k}_2 {\bf k}_3}$, with \cite{Bartolo:2011ee}
\begin{eqnarray}
B_{{\bf k}_1 {\bf k}_2 {\bf k}_3}
&=& 
\frac{6}{5} f_{\rm NL} P(k_1) P(k_2)
 \nonumber \\
 && \left[ 1 + \sum_M \lambda_{2M}
\left( 
 Y_{2M}(\hat{\bf k}_1) + Y_{2M}(\hat{\bf k}_2)  \right) 
\right] \nonumber \\ 
 && 
+ (2~{\rm perm}) ~. \label{eq:zeta3}
\end{eqnarray}
This type of angle dependence is realized in scenarios of anisotropic inflation characterized by the presence of a U(1) gauge field with a small but nonvanishing vacuum expectation value (see e.g., Refs.~\cite{Bartolo:2011ee, Bartolo:2012sd,Bartolo:2015dga} and the reviews~\cite{Dimastrogiovanni:2010sm,Maleknejad:2012fw} for other possibilities).~\footnote{The quadrupolar signature~\eqref{eq:zeta3} is a generic common prediction of inflation models where rotational-invariance breaking is induced by vector fields. In addition, depending on the specific model considered, further contributions to the anisotropic bispectrum
are present; see, e.g., Refs.~\cite{Bartolo:2011ee,Bartolo:2015dga}.}
One can expect these models to leave signatures in the $\mu$-$T$ correlation function, since the isotropic part is of the local type (thus peaked on squeezed triangles), and the $\mu$-$T$ correlation is 
 indeed sensitive to the squeezed limit of the curvature bispectrum ($k_1 \approx k_2 \gg k_3$). 

\begin{figure}[t]
  \begin{tabular}{c}
    \begin{minipage}{1.0\hsize}
  \begin{center}
    \includegraphics[width = 1.0\textwidth]{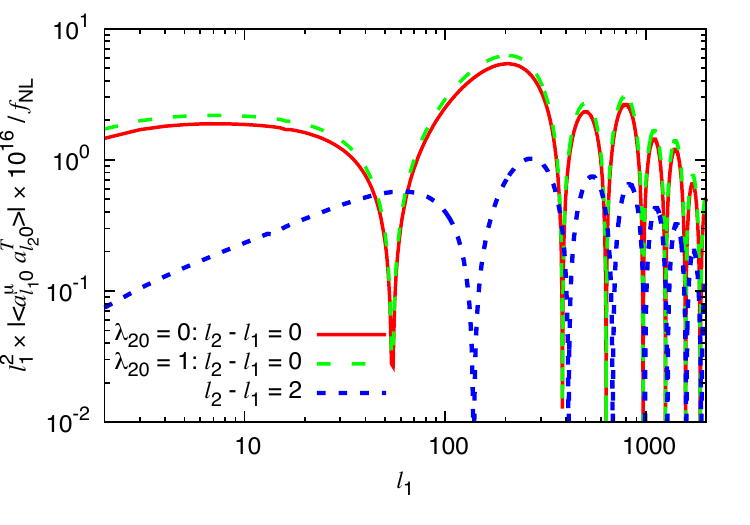}
  \end{center}
\end{minipage}
  \end{tabular}
  \caption{Absolute values of the $\mu$-$T$ power spectra (normalized by $f_{\rm NL}$) for $\lambda_{20} = 0$ and $\ell_1 = \ell_2$ (red solid), $\lambda_{20} = 1$ and $\ell_1 = \ell_2$ (green dashed), and $\lambda_{20} = 1$ and $\ell_1 = \ell_2 - 2$ (blue dotted), with $m_1 = m_2 = 0$ and $\lambda_{2, \pm 1} = \lambda_{2, \pm 2} = 0$. The red solid line is in agreement with a $\mu$-$T$ correlation generated from the usual isotropic local-type NG \cite{Ganc:2012ae}.} \label{fig:Cl_FNL1_Lam2M1_m0}
\end{figure}

 The projection of $\mu({\bf x})$ [Eq.~\eqref{eq:mu}] on the last scattering surface leads to
 \begin{eqnarray}
   a_{\ell m}^\mu &=& \int d^2 \hat{\bf n} \, \mu(x_{\rm ls} \hat{\bf n}) Y_{\ell m}^*(\hat{\bf n}) \nonumber \\ 
   &=& 4\pi (-i)^{\ell} \left[ \prod_{n=1}^2 
\int \frac{d^3 {\bf k}_n}{(2\pi)^3} \zeta_{{\bf k}_n}
\right]
\int d^3 {\bf k}_3 \delta^{(3)}\left(\sum_{n=1}^3 {\bf k}_n \right) \nonumber \\ 
&&Y_{\ell m}^*(\hat{\bf k}_3)
 j_{\ell}(k_3 x_{\rm ls}) f(k_1, k_2, k_3) ~, \label{eq:alm_mu_iso}
 \end{eqnarray}
 where $x \equiv |{\bf x}|$, $\hat{\bf n} \equiv {\bf x} / x$, and $x_{\rm ls}$ is the conformal distance to the last scattering surface. 
 The correlation between $a_{\ell m}^\mu$ and the CMB temperature anisotropy $a_{\ell m}^{T} = 4\pi i^{\ell} \int \frac{d^3 {\bf k}}{(2\pi)^{3}} {\cal T}_{\ell}(k) \zeta_{\bf k} Y_{\ell m}^*(\hat{\bf k})$ [${\cal T}_{\ell}(k)$ being the CMB radiation transfer function]  reads
 \begin{eqnarray}
&& \Braket{a_{\ell_1 m_1}^\mu a_{\ell_2 m_2}^T}
= i^{\ell_2 - \ell_1}  
\left[\prod_{n=1}^2 \int \frac{d^3 {\bf k}_n}{2\pi^2} 
\right] \nonumber \\ 
&&\quad\times  \int d^3 {\bf k}_3 \delta^{(3)}\left(\sum_{n=1}^3 {\bf k}_n \right)
Y_{\ell_1 m_1}^*(\hat{\bf k}_3) Y_{\ell_2 m_2}^*(\hat{\bf k}_3) 
\nonumber \\ 
&&\quad\times  {\cal T}_{\ell_2}(k_3)  j_{\ell_1}(k_3 x_{\rm ls})
          f(k_1, k_2, k_3) B_{{\bf k}_1 {\bf k}_2 {\bf k}_3}  \, .
   \end{eqnarray}
 The fact that $f(k_1, k_2, k_3) \times B_{{\bf k}_1 {\bf k}_2 {\bf k}_3} $ filters the squeezed signals $k_1 \approx k_2 \gg k_3$ reduces this to
\begin{eqnarray}
&&\Braket{a_{\ell_1 m_1}^\mu a_{\ell_2 m_2}^T}
\approx 4\pi i^{\ell_2 - \ell_1}
\int \frac{k_1^2 dk_1}{2\pi^2}
\int \frac{d^3 {\bf k}_3}{2\pi^2} \nonumber \\ 
&&\quad\times Y_{\ell_1 m_1}^*(\hat{\bf k}_3) Y_{\ell_2 m_2}^*(\hat{\bf k}_3) 
{\cal T}_{\ell_2}(k_3)  j_{\ell_1}(k_3 x_{\rm ls}) f(k_1, k_1, k_3) \nonumber \\ 
&&\quad\times   \frac{12}{5} f_{\rm NL} P(k_1) P(k_3)
 \left[ 1 + \sum_M \lambda_{2M} Y_{2M}(\hat{\bf k}_3)  
\right]~.
\end{eqnarray}
Here, the contribution from $\sum_M \lambda_{2M} Y_{2M}(\hat{\bf k}_1)$ has vanished, since $\int d^2 \hat{\bf k}_1 Y_{2M}(\hat{\bf k}_1) = 0$.  
The $\hat{\bf k}_3$ integrals can be analytically computed, reading $\int d^2 \hat{\bf k}_3  Y_{\ell_1 m_1}^*(\hat{\bf k}_3) Y_{\ell_2 m_2}^*(\hat{\bf k}_3) = (-1)^{m_1} \delta_{\ell_1, \ell_2} \delta_{m_1, -m_2}$ and
\begin{eqnarray}
 &&  \int d^2 \hat{\bf k}_3  Y_{\ell_1 m_1}^*(\hat{\bf k}_3) Y_{\ell_2 m_2}^*(\hat{\bf k}_3) Y_{2M}(\hat{\bf k}_3) \nonumber \\
  &&\qquad = (-1)^{m_1 + m_2}
  h_{\ell_1 \ell_2 2}
  \left(
  \begin{array}{ccc}
  \ell_1 & \ell_2 & 2 \\
  -m_1 & -m_2 & M 
  \end{array}
 \right)
    ~,
\end{eqnarray}
where we have introduced a geometrical function:
\begin{eqnarray}
h_{l_1 l_2 l_3} 
\equiv \sqrt{\frac{(2 l_1 + 1)(2 l_2 + 1)(2 l_3 + 1)}{4 \pi}}
\left(
  \begin{array}{ccc}
  l_1 & l_2 & l_3 \\
   0 & 0 & 0
  \end{array}
 \right), 
\end{eqnarray}
which vanishes for configurations that do not respect the triangle inequality, $|l_1 - l_2| \leq l_3 \leq l_1 + l_2$, and parity symmetry, $l_1 + l_2 + l_3 = {\rm even}$. A final form is written as  
\begin{eqnarray}
\Braket{a_{\ell_1 m_1}^{\mu} a_{\ell_2 m_2}^{T}}
&=&\bar{C}_{\ell_1}^{\mu T} (-1)^{m_1} \delta_{\ell_1, \ell_2} \delta_{m_1, -m_2} \nonumber \\
&& + (-1)^{m_2} C_{\ell_1 m_1, \ell_2 -m_2}^{\mu T}\, ,
\end{eqnarray}
where 
\begin{eqnarray}
\bar{C}_\ell^{\mu T}
&=& f_{\rm NL} G_{\ell \ell}^{\mu T}
  ~, \\
C_{\ell_1 m_1 \ell_2 m_2}^{\mu T} &=& 
 i^{\ell_2 - \ell_1} 
f_{\rm NL} G_{\ell_1 \ell_2}^{\mu T}
 (-1)^{m_1}
h_{\ell_1 \ell_2 2} \nonumber \\  
&&  \sum_M \lambda_{2M}
\left(
  \begin{array}{ccc}
  \ell_1 & \ell_2 & 2 \\
  -m_1 & m_2 & M 
  \end{array}
 \right) ~, \label{eq:muT_ani_lambda} \\
G_{\ell_1 \ell_2}^{\mu T} &\equiv& 
4\pi
\int \frac{k_1^2 dk_1 }{2\pi^2}
\int \frac{k_3^2 dk_3 }{2\pi^2}
 {\cal T}_{\ell_2}(k_3)  j_{\ell_1}(k_3 x_{\rm ls}) \nonumber \\ 
&& f(k_1, k_1, k_3) \frac{12}{5} P(k_1)P(k_3) ~.
\end{eqnarray}
One can find that, via the angular integrals, the quadrupolar angular dependence in Eq.~\eqref{eq:zeta3} produces $h_{\ell_1 \ell_2 2}$ in Eq.~\eqref{eq:muT_ani_lambda}, resulting in {\em nonvanishing off-diagonal components} when 
$|\ell_1 - \ell_2| = 2$, as well as diagonal ones: $\ell_1 = \ell_2$. Figure~\ref{fig:Cl_FNL1_Lam2M1_m0} displays a numerical evaluation of $\Braket{a_{\ell_1 m_1}^{\mu} a_{\ell_2 m_2}^{T}}$. One can see from this figure that nonzero $\lambda_{2M}$ creates both diagonal (green dashed line) and off-diagonal (blue dotted line) signals with different shapes. We will study the detectability of $\lambda_{2M}$ in Sec.~\ref{sec:Fisher}.

The primordial models generating Eq.~\eqref{eq:zeta3} often produce a quadrupolar asymmetry also in the curvature power spectrum (see, e.g., Ref.~\cite{Bartolo:2012sd} for a specific inflation model where there is a coupling between the inflaton field and a vector kinetic term $F^2$, and see Refs.~\cite{Dimastrogiovanni:2010sm,Maleknejad:2012fw} for reviews). The amplitude of the quadrupolar asymmetry in the curvature power spectrum  is generally characterized via the parameter $g_*$~\cite{Ackerman:2007nb}.
One could expect some signature of this modulation to be imprinted in a $\mu$-$\mu$ correlation, since we know that the latter is sensitive to the $\tau_{\rm NL}$ trispectrum parameter (see Ref.~\cite{Pajer:2012vz}), 
which in turn describes large scales modulations of small-scale power \cite{Pearson:2012ba, Ade:2013ydc}. Of course, a standard isotropic primordial power spectrum is also expected to produce a nonvanishing $\mu$-$\mu$ correlation, via its disconnected contribution to the trispectrum. The hope here is that the anisotropic nature of the $g_*$ signal can again produce off-diagonal couplings, 
that are not sourced by the isotropic part. However, this is not the case, as we now explicitly show. 

  We start with a power spectrum in a general anisotropic form: $\Braket{\zeta_{{\bf k}_1} \zeta_{{\bf k}_2}} = (2\pi)^3 \delta^{(3)}({\bf k}_1 + {\bf k}_2) P({\bf k}_1)$. In this form, the $g_*$ parametrization is given as $P({\bf k}) = P(k)[1 + g_* (\hat{\bf k} \cdot \hat{\bf p})^2]$. We are thus now interested in the study of $\mu$-$\mu$ correlations arising from the above anisotropic primordial power spectrum:
\begin{eqnarray}
&& \Braket{\prod_{n=1}^2 a_{\ell_n m_n}^\mu}
  = 2(-i)^{\ell_1 + \ell_2} \left[ \prod_{n=1}^2 \int \frac{d^3 {\bf k}_n}{2\pi^2}
    P({\bf k}_n)   \right] 
\nonumber \\
&&\quad \times
\int d^3 {\bf k}_3
\delta^{(3)}\left(\sum_{n=1}^3 {\bf k}_n\right) 
Y_{\ell_1 m_1}^*(\hat{\bf k}_3) Y_{\ell_2 m_2}^*(-\hat{\bf k}_3) \nonumber \\ 
&&\quad \times j_{\ell_1}(k_3 x_{\rm ls}) j_{\ell_2}(k_3 x_{\rm ls}) f^2(k_1, k_2, k_3) 
~. 
\end{eqnarray}
The integrand is strongly peaked on squeezed configurations ($k_1 \approx k_2 \gg k_3$) due to the filtering by $f(k_1, k_2, k_3)$. The $\hat{\bf k}_3$ integral can then be reduced to $\int d^2 \hat{\bf k}_3 Y_{\ell_1 m_1}^*(\hat{\bf k}_3) Y_{\ell_2 m_2}^*(-\hat{\bf k}_3)$, and this results in vanishing off-diagonal components due to usual orthonormality of the spherical harmonics. 
We thus arrive at the conclusion that, in the $\mu$-$\mu$ power spectrum, the anisotropic signatures of the curvature power spectrum cannot be differentiated from the isotropic ones.

  On the other hand, a connected anisotropic contribution to the curvature trispectrum is expected to create some non-vanishing off-diagonal signals, as well as diagonal ones due to $\tau_{\rm NL}$. It is known that anisotropic trispectrum signals can also be produced in vector inflation models \cite{Bartolo:2009kg, ValenzuelaToledo:2009nq, Rodriguez:2013cj, Shiraishi:2013vja, Abolhasani:2013zya}, while there may exist other possible models where unknown anisotropic signals are realized. The trispectrum parametrization is still nontrivial and the resultant $\mu$-$\mu$ signatures need to be analyzed on a case-by-case basis. We leave this for future work.

\section{$\mu$-$\mu$ correlation due to dipolar power asymmetry}\label{sec:dipolarpower}

We now consider a case in which primordial curvature perturbations depend (mildly) on a spatial position ${\bf x}$ \cite{Dai:2013kfa, Lyth:2013vha, Jazayeri:2014nya}: 
\begin{eqnarray}
\zeta_{\bf k}({\bf x}) &=& \zeta_{\bf k}
\left[1 + A(k) \frac{{\bf x} \cdot \hat{\bf p}}{x_{\rm ls}} \right] \, ,\label{eq:zeta_def_dipole}
\end{eqnarray}
where $\hat{\bf p}$ is the direction of a dipolar modulation of CMB power. Models which lead to this type of ansatz for the primordial power spectrum have been proposed to explain the observed power asymmetry in the CMB temperature sky \cite{Ade:2013nlj, Ade:2015sjc, Hansen:2004vq, Gordon:2005ai, Eriksen:2007pc, Gordon:2006ag, Akrami:2014eta, Jazayeri:2014nya, Aiola:2015rqa}. Again, the dipolar modulation of power motivates the search for off-diagonal signatures in the $\mu$ autocorrelation function [note that the correlation between the $\mu$ distortion and the temperature CMB anisotropy vanishes in this case, since $\mu$-$T$ is sensitive to the bispectrum, and we assume Gaussianity of the curvature perturbation in Eq.~\eqref{eq:zeta_def_dipole}].

One can expand the dipolar modulation in spherical harmonics, $Y_{1M}$, as
\begin{eqnarray}
\zeta_{\bf k}({\bf x}) &=& \zeta_{\bf k}
\left[1 + \sum_M A_{1M}(k) Y_{1M}(\hat{\bf n}) \right] ~, \label{eq:zeta_def_dipole_A1M} \\
A_{1M}(k)  &=& \frac{4\pi}{3} A(k) \frac{x}{x_{\rm ls}} Y_{1M}^*(\hat{\bf p})
~.
\end{eqnarray}
 We now discuss the $\mu$-$\mu$ power spectrum in a toy model in which $A(k) = const$. Detailed derivation of the $\mu$-$\mu$ correlator induced by a general $A(k)$ is presented in Appendix~\ref{appen:dipolarpowerfull}. 
Substituting Eq.~\eqref{eq:zeta_def_dipole_A1M} into Eq.~\eqref{eq:mu} and performing the spherical harmonic transform of $\mu$,  
projected on the last scattering surface, $a_{\ell m}^\mu = \int d^2 \hat{\bf n} \mu(x_{\rm ls} \hat{\bf n}) Y_{\ell m}^*(\hat{\bf n})$, we obtain
\begin{eqnarray}
a_{\ell m}^\mu 
&=& \bar{a}_{\ell m}^{\mu} + 2 (-1)^{m} \sum_{L M}  \bar{a}_{L M}^{\mu}
\nonumber \\ 
&&\times 
\sum_{M'} A_{1M'}
h_{\ell L 1}
\left(
  \begin{array}{ccc}
  \ell & L & 1 \\
   -m & M & M' 
  \end{array}
 \right) ~, \label{eq:alm_nu}
\end{eqnarray}
where we have ignored a ${\cal O}(A^2)$ term because of the smallness of $A$. One can notice that, in $\ell$ space, the term including the dipolar asymmetry gets convolved with the isotropic part $\bar{a}_{\ell m}^{\mu}$, which is expressed in the same form as Eq.~\eqref{eq:alm_mu_iso}. 
Evaluating the autocorrelation in the squeezed limit yields 
\begin{eqnarray}
\Braket{a_{\ell_1 m_1}^{\mu} a_{\ell_2 m_2}^{\mu}}
&=& \bar{C}_{\ell_1}^{\mu \mu} (-1)^{m_1} \delta_{\ell_1, \ell_2} \delta_{m_1, -m_2} \nonumber \\
&&+ (-1)^{m_2} C_{\ell_1 m_1, \ell_2 -m_2}^{\mu\mu}\, ,
\end{eqnarray}
where 
\begin{eqnarray}
\bar{C}_{\ell}^{\mu \mu} &\approx&  
8\pi \int \frac{k_1^2 dk_1}{2 \pi^2} \int \frac{k_3^2 dk_3}{2 \pi^2} \nonumber \\ 
&& j_{\ell}^2(k_3 x_{\rm ls}) f^2(k_1, k_1, k_3) P^2(k_1) ~, \label{eq:mumu_iso} \\
C_{\ell_1 m_1, \ell_2 m_2}^{\mu\mu} 
&\approx& 2  (-1)^{m_1} \left( \bar{C}_{\ell_1}^{\mu \mu} + \bar{C}_{\ell_2}^{\mu \mu} \right)
h_{\ell_1 \ell_2 1} \nonumber \\
&& \sum_M A_{1M}
\left(
  \begin{array}{ccc}
  \ell_1 & \ell_2 & 1 \\
   -m_1 & m_2 & M 
  \end{array}
 \right) ~, \label{eq:mumu_ani_A}
\end{eqnarray}
with $P(k)$ denoting the isotropic component of the curvature power spectrum, defined as $\Braket{\zeta_{{\bf k}_1} \zeta_{{\bf k}_2}} = (2\pi)^3 \delta^{(3)}({\bf k}_1 + {\bf k}_2) P(k_1)$.

The result above tells us that a nonzero dipolar asymmetry creates a distinctive off-diagonal signature, satisfying $|\ell_1 - \ell_2| = 1$, in the $\mu$-$\mu$ power spectrum. This type of coupling 
is expected, given the dipolar nature of the modulation. The same type of signature, for this class of models, is seen (on different scales) in the CMB temperature power spectrum \cite{Hanson:2009gu}:
\begin{eqnarray}
C_{\ell_1 m_1, \ell_2 m_2}^{TT} 
&=& (-1)^{m_1}
 \left( \bar{C}_{\ell_1}^{TT} + \bar{C}_{\ell_2}^{TT} \right)
    h_{\ell_1 \ell_2 1} \nonumber \\ 
&&\sum_{M} A_{1M} 
\left(
  \begin{array}{ccc}
  \ell_1 & \ell_2 & 1 \\
   -m_1 & m_2 & M 
  \end{array}
 \right) ~. \label{eq:TT_ani_A}
\end{eqnarray}
Note that $\bar{C}_\ell^{\mu \mu}$ behaves as a white noise term, since this is sourced by a Gaussian isotropic curvature perturbation, the value of which is evaluated 
as $\bar{C}_\ell^{\mu \mu} \lesssim 10^{-28} $ \cite{Pajer:2012vz}.

We know that the dipolar modulation observed in the CMB power spectrum affects only relatively large scales, $\ell \lesssim 60$ (see e.g., Refs.~\cite{Hoftuft:2009rq, Ade:2013nlj, Akrami:2014eta, Ade:2015sjc, Aiola:2015rqa}). 
The choice $A(k) = const$, that we just made above, is thus unable to explain current observations. A red-tilted $k$ dependence must be introduced in $A(k)$, in such a way as to make the modulation 
 negligible for $k \gtrsim 60 / x_{\rm ls} = 0.004 ~ {\rm Mpc}^{-1}$. In this case one can see that the $k_1$ integral in formula \eqref{eq:mumu_ani_any_A} vanishes, since the window function $f(k_1,k_1,k_3)$ is not suppressed only for 
 large $k_1$ ($k_1 > k_D(z_f) \approx 46 ~ {\rm Mpc}^{-1}$, where $k_D$ is the Silk damping scale). It is then not possible to use $\mu$-$\mu$ to test the possible primordial origin of the observed temperature asymmetry. This reflects the fact that $\mu$ CMB distortions probe very small scales. On the other 
hand, {we used the toy model~\eqref{eq:zeta_def_dipole}} to show that the off-diagonal coupling pointed out above could still be used to build observational tests of rotational invariance at otherwise unaccessible CMB scales.

\section{Fisher forecast}\label{sec:Fisher}

We now evaluate the detectability of the signal discussed above through a simple Fisher matrix analysis. In the limit that asymmetries are regarded as tiny modulations of the isotropic part, 
a minimum-variance estimator for the magnitudes of the asymmetry (in our paper, ${\bf h} = A_{1 M}$ or $ \lambda_{2 M}$) can be written as \cite{Hanson:2009gu, Hanson:2010gu}
\begin{eqnarray}
\hat{\bf h} = \frac{1}{2} {\bf F}^{-1}
 \tilde{\bf a}^\dagger
\frac{\delta {\bf C}}{\delta {\bf h}^\dagger} 
 \tilde{\bf a} ~,
\end{eqnarray}
where $\tilde{\bf a} \equiv {\bf C}_{\rm obs}^{-1} {\bf a}$ denotes the observed harmonic coefficients after application of inverse-variance filtering. In the following discussion, we assume that off-diagonal components of the inverse of the covariance matrix are negligibly small; thus the Fisher matrix can be diagonalized. 

The variance term ${\bf C}_{\rm obs}$ is composed of cosmic variance and instrumental noise. In what follows, we will first investigate an ideal case where ${\bf C}_{\rm obs}^{\mu \mu}$ is determined by cosmic variance alone, and 
then some cases involving instrumental noise levels similar to those of {\it Planck} \cite{Planck:2006aa}, PIXIE \cite{Kogut:2011xw} and CMBpol \cite{Baumann:2008aq}. For $\mu$-$\mu$ noise spectra, we assume $N_\ell^{\mu \mu} = N_\mu \exp\left(\ell^2 / \ell_\mu^2 \right)$, with $(N_\mu, \ell_\mu) = (10^{-15}, 861)$ ({\it Planck}), $(10^{-17}, 84)$ (PIXIE) and $(2\times 10^{-18}, 1000)$ (CMBpol) \cite{Ganc:2012ae,Ganc:2014wia}. On the other hand, we know that CMB temperature measurements on large scales, which are the only ones to give contributions to the $\mu$-$T$ correlation signal, 
are completely cosmic-variance dominated. Thus, we can ignore instrumental noise in ${\bf C}_{\rm obs}^{TT}$, so that $C_{\ell, \rm obs}^{TT} \approx \bar{C}_\ell^{TT}$.

\subsection{Anisotropic non-Gaussianity}

Assuming $(C_{\ell, \rm obs}^{\mu T})^2 \ll C_{\ell, \rm obs}^{TT} C_{\ell, \rm obs}^{\mu \mu}$, the Fisher matrix for $\lambda_{2 M}$ becomes
\begin{eqnarray}
F_{\lambda}^{(\mu T)} = \frac{f_{\rm NL}^2}{5}
 \sum_{\ell_1, \ell_2 = 2}^{\ell_{\rm max}} h_{\ell_1 \ell_2 2}^2 
\frac{\left(G_{\ell_1 \ell_2}^{\mu T}\right)^2}
{C_{\ell_1, \rm obs}^{\mu\mu} C_{\ell_2, \rm obs}^{TT}} ~. \label{eq:F_muT_lambda}
\end{eqnarray}

\begin{figure}[t]
  \begin{tabular}{c}
    \begin{minipage}{1.0\hsize}
  \begin{center}
    \includegraphics[width = 1.0\textwidth]{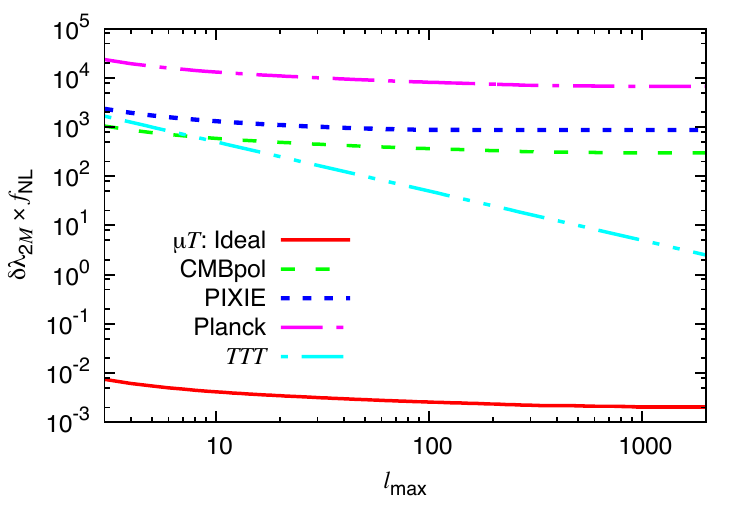}
  \end{center}
\end{minipage}
\end{tabular}
  \caption{Expected $1\sigma$ errors on $\lambda_{2M}$ for $f_{\rm NL} = 1$ estimated from the $\mu$-$T$ correlation in the noise-free ideal case and in the cases including the instrumental noises of {\it Planck}, PIXIE and CMBpol. For comparison, the error estimated from the temperature bispectrum \cite{Bartolo:2011ee} is also plotted. Following previous results in the literature, we take the cosmic variance level as $\bar{C}_\ell^{\mu \mu} = 10^{-28}$.} \label{fig:dFNL_dLam2M_FNL1}
\end{figure}

At first, we focus on a noise-free, cosmic-variance-dominated ideal case, leading to $C_{\ell, \rm obs}^{\mu\mu} = \bar{C}_{\ell}^{\mu\mu} = const$. The scaling is evaluated as follows. Since $h_{\ell_1 \ell_2 2}$ vanishes except for $\ell_2 = \ell_1, \ell_1 \pm 2$, we may drop $\sum_{\ell_2}$ by replacing $\ell_2$ with $\ell_1$. In the Sachs-Wolfe limit ${\cal T}_\ell(k) \to -j_\ell(k x_{\rm ls}) / 5$, and the scale-invariant power spectrum $P(k) \propto k^{-3}$ yields $G_{\ell_1 \ell_1}^{\mu T} \propto \bar{C}_{\ell_1}^{TT} \propto \ell_1^{-2}$. These estimates, and $h_{\ell_1 \ell_1 2}^2 \propto \ell_1$, lead to 
\begin{eqnarray}
F_{\lambda}^{(\mu T \rm CV)} \propto 
\sum_{\ell_1}^{\ell_{\rm max}} 
\ell_1  \frac{(\ell_1^{-2})^2}
{\ell_1^{-2}} \propto \ln \ell_{\rm max}  \, ,
\end{eqnarray}
where CV stands for cosmic-variance dominated. We have checked that this scaling is in agreement with the full numerical evaluation of Eq.~\eqref{eq:F_muT_lambda} in the ideal case (corresponding to red solid line in Fig.~\ref{fig:dFNL_dLam2M_FNL1}). Combining the numerical results with this scaling, we finally obtain an expected $1\sigma$ error bar $1 / \sqrt{F}$ as
\begin{eqnarray}
\delta \lambda_{2M}^{(\mu T \rm CV)} 
\approx \frac{5 \cdot 10^{-3}}{f_{\rm NL}} 
\left(\frac{\bar{C}_\ell^{\mu \mu}}{10^{-28}}\right)^{1/2} (\ln \ell_{\rm max})^{-1/2}~.
\end{eqnarray}

On the other hand, an expected error obtained from the CMB temperature bispectrum has a different amplitude and scaling of $\ell_{\rm max}$ as $\delta \lambda_{2M}^{(TTT)} \approx (2.5 / f_{\rm NL}) (2000/\ell_{\rm max})$ \cite{Bartolo:2011ee}. Comparison of these results indicates that the $\mu$-$T$ correlation can be a much more powerful observable, in case we can remove instrumental uncertainties completely (see Fig.~\ref{fig:dFNL_dLam2M_FNL1}).

The expected error bars in the {\it Planck}-like, PIXIE-like and CMBpol-like surveys are obtained by computing Eq.~\eqref{eq:F_muT_lambda} with $C_{\ell, \rm obs}^{\mu \mu} = \bar{C}_\ell^{\mu\mu} + N_\ell^{\mu\mu}$. The numerical results depicted in Fig.~\ref{fig:dFNL_dLam2M_FNL1} show that, even for potential future surveys like CMBpol, the $\mu$-$T$ constraint is not competitive with temperature bispectrum measurements. However, an ideal noise-free experiment can achieve extremely tight constraints. 
This, as usual for this type of signal, is due to the fact that $\mu$ spectral distortions provide (integrated) information up to very high $\ell$. Although achievable only with futuristic surveys and very low noise levels, the potential of this kind of tests is thus large.

\subsection{Dipolar power asymmetry}

When we consider our toy model of dipolar asymmetry and estimate $A_{1 M} (= const)$ from the $\mu$-$\mu$ power spectrum, the Fisher matrix is written as 
\begin{eqnarray}
F_A^{(\mu \mu)} = 
 \frac{2}{3} \sum_{\ell_1, \ell_2 = 2}^{\ell_{\rm max}}   
h_{\ell_1 \ell_2 1}^2
 \frac{\left( \bar{C}_{\ell_1}^{\mu\mu} + \bar{C}_{\ell_2}^{\mu\mu} \right)^2}{C_{\ell_1, \rm obs}^{\mu\mu } C_{\ell_2, \rm obs}^{\mu\mu}}
~. \label{eq:F_mu_A}
\end{eqnarray}

In the noise-free ideal case, we can write $C_{\ell, \rm obs}^{\mu\mu} = \bar{C}_\ell^{\mu \mu} = const$. This reduces the Fisher matrix to  
\begin{eqnarray}
F_A^{(\mu \mu \rm CV)} &=& 
 \frac{8}{3} \sum_{\ell_1, \ell_2 = 2}^{\ell_{\rm max}} h_{\ell_1 \ell_2 1}^2 \nonumber \\
&=& \frac{2}{\pi} (\ell_{\rm max} + 3)  (\ell_{\rm max} - 2)~, \label{eq:F_mu_A_CV}
\end{eqnarray} 
and hence an expected $1\sigma$ error bar $1 / \sqrt{F}$ is estimated as 
\begin{eqnarray}
\delta A_{1M}^{(\mu\mu \rm CV)} \approx \frac{1.3}{ \ell_{\rm max}}~.
\end{eqnarray}
This can be compared to the results estimated from the temperature power spectrum. The Fisher matrix $F_A^{(TT)}$ becomes the same form of $F_A^{(\mu \mu)}$, aside from a prefactor 4 due to the difference of factor 2 between $C_{\ell_1 m_1, \ell_2 m_2}^{\mu\mu} $ in  Eq.~\eqref{eq:mumu_ani_A} and $C_{\ell_1 m_1, \ell_2 m_2}^{TT} $ in Eq.~\eqref{eq:TT_ani_A}. Since $\bar{C}_{\ell}^{TT}$ and $\bar{C}_{\ell \pm 1}^{TT}$ take almost identical values, the Fisher matrix for the ideal cosmic-variance-limited case can be simplified like Eq.~\eqref{eq:F_mu_A_CV}, and we finally find $\delta A_{1M}^{(TT \rm CV)} \approx 2 \delta A_{1M}^{(\mu\mu \rm CV)}$.

\begin{figure}[t]
  \begin{tabular}{c}
    \begin{minipage}{1.0\hsize}
  \begin{center}
    \includegraphics[width = 1.0\textwidth]{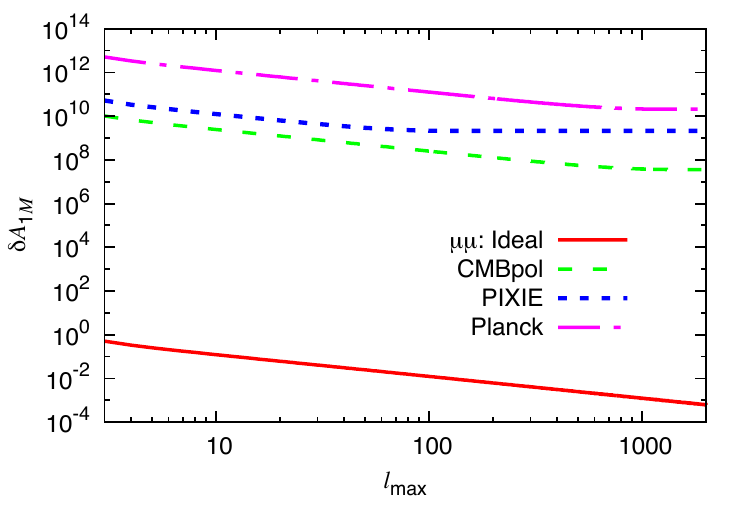}
  \end{center}
\end{minipage}
\end{tabular}
  \caption{Expected $1\sigma$ errors on $A_{1M}$ estimated from the $\mu$-$\mu$ correlation in the noise-free ideal case and in the cases including instrumental noise levels of {\it Planck}, PIXIE and CMBpol. We here adopt $\bar{C}_\ell^{\mu \mu} = 10^{-28}$.}  \label{fig:dA1M}
\end{figure}

The errors for the cases with more realistic noise levels are computed with $C_{\ell, \rm obs}^{\mu \mu} = \bar{C}_\ell^{\mu\mu} + N_\ell^{\mu\mu}$. The results plotted in Fig.~\ref{fig:dA1M} are consistent with the expectation from Eq.~\eqref{eq:F_mu_A} that, for $\ell_{\rm max} \lesssim \ell_\mu$, the error bars simply get bigger by a factor $N_\mu / \bar{C}_\ell^{\mu \mu}$ with respect to the noise-free ideal case. We stress again that we are considering 
here just a toy model with a constant modulation extending up to very high $\ell$. The dipolar asymmetry that is found in the CMB extends up to at most $\ell \sim 100$, and it thus leaves no signature in $\mu$-$\mu$.

Before concluding this section, we would like to stress that the forecasts presented here do not take into account important issues, like foreground subtraction. Such issues play, of course, a crucial role in actual measurements, and would most likely affect the expected error bars, plotted in Figs.~\ref{fig:dFNL_dLam2M_FNL1} and \ref{fig:dA1M}, in a survey-dependent way (with e.g. PIXIE expected to be significantly more efficient than {\it Planck} or CMBpol at component separation for $\mu$ reconstruction). An accurate investigation of these effects is beyond the scope of this simple analysis. Our main goal in this section was just to show that, while interesting information on isotropy-breaking signatures {\em is} present in the $\mu$-$T$ and $\mu$-$\mu$ correlations, only futuristic surveys will be able to extract it (as is the case also for many other NG signatures, such as primordial NG of the local type, or signatures induced by primordial magnetic fields).

\section{Conclusions}\label{sec:conclusions}

We studied CMB $\mu$ spectral distortion signatures arising from dissipation of acoustic waves in models characterized by 
primordial breaking of rotational invariance. We find that such anisotropic primordial scenarios predict distinctive off-diagonal 
signatures in the $\mu$-$T$ correlation matrix.

More specifically, for NG models generating bispectra with a quadrupolar asymmetry, as generally predicted by anisotropic inflationary scenarios, we find nonvanishing $\mu$-$T$ correlations with a distinct $|\ell_1 - \ell_2| = 2$ signature. The coupling we find is consistent with what we intuitively expected, due to the quadrupolar nature of the rotation-invariance-breaking terms, and since these bispectra peak in the squeezed limit (a necessary condition to source $\mu$-$T$ correlations). For these scenarios, we also check that anisotropic power spectrum modulations, characterized by the $g_*$ parameter~\cite{Ackerman:2007nb}, do not leave any specific off-diagonal signature in the $\mu$-$\mu$ correlation function.

These signals are not easy to observe, and will not fall into the detectability 
range for next-generation missions, such as PIXIE, or even for proposed future surveys like CMBpol. However, our Fisher matrix analysis for an ideal cosmic-variance-dominated experiment shows that spectral-distortion-based measurements can in principle be much more powerful than temperature-only estimates 
if noise is negligible up to small enough scales. This, as originally pointed out in Ref.~\cite{Pajer:2012vz} for local bispectrum and trispectrum measurements, is essentially due to the fact that $\mu$ distortions are sensitive to primordial fluctuations up to very large wave numbers, where temperature anisotropies 
have been erased by Silk damping. It is therefore clear that, although advanced futuristic surveys with very high sensitivity are necessary in order to study the effects mentioned here, 
the information content of such effects is very large, and worth pointing out.

As a further example, besides anisotropic inflationary bispectra, we also studied Gaussian toy models with a dipolar asymmetry in the primordial power spectrum, extending up 
to very small scales. In this case, a trispectrum signature is seen in the $\mu$-$\mu$ correlation function, as a coupling between 
different $\ell_1$ and $\ell_2$, with $|\ell_1 - \ell_2| = 1$. These models clearly do not reflect the observed CMB dipolar asymmetry, which extends only 
to relatively small scales, and does not leave any imprint on $\mu$-$\mu$. However, they show that $\mu$ distortions allow one in principle 
to test statistical isotropy on scales much smaller than what is actually probed by CMB temperature and polarization data.

\acknowledgments
We thank Frode Hansen for very useful discussions about the CMB hemispherical power asymmetry. MS was supported in part by a Grant-in-Aid for JSPS Research under Grants No.~25-573 and No.~27-10917, and in part by the World Premier International Research Center Initiative (WPI Initiative), MEXT, Japan. This work was supported in part by ASI/INAF Agreement I/072/09/0 for the Planck LFI Activity of Phase E2.

%

\appendix

\section{Derivation of $\mu$-$\mu$ correlator generated from dipolar power asymmetry} \label{appen:dipolarpowerfull}

We here present the details of the computations on the $\mu$-$\mu$ power spectrum induced by the dipolar modulation \eqref{eq:zeta_def_dipole} or \eqref{eq:zeta_def_dipole_A1M}. A formula obtained here is applicable to $A(k)$ with any scale dependence.

A spherical harmonic transformation of $\mu({\bf x})$ [Eq.~\eqref{eq:mu}] leads to
\begin{eqnarray}
  a_{\ell m}^\mu &=&
\left[ \prod_{n=1}^2 
\int \frac{d^3 {\bf k}_n}{(2\pi)^3} \zeta_{{\bf k}_n}
\right] 
\int d^3 {\bf k}_3 
 \delta^{(3)}\left(\sum_{n=1}^3 {\bf k}_n\right) \nonumber \\ 
 && f(k_1, k_2, k_3) 
 \int d^2 \hat{\bf n}
Y_{\ell m}^*(\hat{\bf n})
e^{-i {\bf k}_3 \cdot x_{\rm ls} \hat{\bf n}} \nonumber \\ 
&& \left(1 + \sum_{M'} [A_{1M'}(k_1) + A_{1M'}(k_2)] Y_{1M'}(\hat{\bf n}) \right),
\end{eqnarray}
where we have dropped the ${\cal O}(A^2)$ term. With an identity
\begin{eqnarray}
e^{-i {\bf k}_3 \cdot x_{\rm ls} \hat{\bf n}} = \sum_{LM} 4\pi (-i)^L j_L(k_3 x_{\rm ls}) Y_{LM}^*(\hat{\bf k}_3) Y_{LM}(\hat{\bf n}),  
\end{eqnarray}
the above $\hat{\bf n}$ integrals are computed as
$\int d^2 \hat{\bf n} Y_{\ell m}^*(\hat{\bf n}) Y_{LM}(\hat{\bf n}) = \delta_{\ell, L} \delta_{m, M}$ and 
\begin{eqnarray}
 && \int d^2 \hat{\bf n} 
  Y_{\ell m}^*(\hat{\bf n})
  Y_{LM}(\hat{\bf n})
  Y_{1M'}(\hat{\bf n}) \nonumber \\
  &&\qquad = (-1)^{m} h_{\ell L 1}
  \left(
  \begin{array}{ccc}
    \ell & L & 1 \\
    -m & M & M' 
  \end{array}
  \right) ~,
\end{eqnarray}
and we finally reach $a_{\ell m}^\mu = \bar{a}_{\ell m}^\mu + a_{\ell m}^{\mu (d)}$, 
where 
\begin{eqnarray}
 a_{\ell m}^{\mu (d)} &=& (-1)^m \sum_{LM} 4\pi (-i)^{L} 
\left[ \prod_{n=1}^2 
\int \frac{d^3 {\bf k}_n}{(2\pi)^3} 
\zeta_{{\bf k}_n} \right] \nonumber \\ 
&&
  \int d^3 {\bf k}_3 
 \delta^{(3)}\left(\sum_{n=1}^3 {\bf k}_n\right)  
Y_{L M}^*(\hat{\bf k}_3) j_{L}(k_3 x_{\rm ls})  \nonumber \\ 
&& f(k_1, k_2, k_3)
\sum_{M'}
 \left[A_{1M'}(k_1) + A_{1M'}(k_2) \right] \nonumber \\ 
&& h_{\ell L 1}
\left(
  \begin{array}{ccc}
  \ell & L & 1 \\
   -m & M & M' 
  \end{array}
 \right) ~.
  \end{eqnarray}
One can easily see that, for $A(k) = const$, this recovers Eq.~\eqref{eq:alm_nu}.

The power spectrum modulation due to $A(k)$ is given as
$C_{\ell_1 m_1, \ell_2 m_2}^{\mu\mu} =
\Braket{a_{\ell_1 m_1}^{\mu (d)} \bar{a}_{\ell_2 m_2}^{\mu *} } + 
\Braket{\bar{a}_{\ell_1 m_1}^\mu a_{\ell_2 m_2}^{\mu (d) *} }$. After computing the trispectrum of curvature perturbations, the first term becomes
\begin{eqnarray}
  &&  \Braket{a_{\ell_1 m_1}^{\mu (d)} \bar{a}_{\ell_2 m_2}^{\mu *} }
  = (-1)^{m_1} \sum_{LM} 2 i^{\ell_2}
  (-i)^{L} 
\left[ \prod_{n=1}^2 
\int \frac{d^3 {\bf k}_n}{2 \pi^2}  \right] \nonumber \\ 
&&\quad \times   \int d^3 {\bf k}_3 
 \delta^{(3)}\left(\sum_{n=1}^3 {\bf k}_n\right)  
 Y_{L M}^*(\hat{\bf k}_3) Y_{\ell_2 m_2}(\hat{\bf k}_3)
 \nonumber \\ 
 &&\quad  \times j_{L}(k_3 x_{\rm ls}) j_{\ell_2}(k_3 x_{\rm ls})
 f^2(k_1, k_2, k_3) P(k_1)P(k_2) h_{\ell_1 L 1} \nonumber \\ 
&&\quad  \times \sum_{M'}
\left[A_{1M'}(k_1) + A_{1M'}(k_2) \right]
\left(
  \begin{array}{ccc}
  \ell_1 & L & 1 \\
   -m_1 & M & M' 
  \end{array}
  \right). 
  \end{eqnarray}
Then, the fact that the transfer function $f(k_1, k_2, k_3)$ filters the squeezed-limit signals $k_1 \approx k_2 \gg k_3 $ simplifies the $\hat{\bf k}_3$ integral as $\int d^2 \hat{\bf k}_3 
 Y_{L M}^*(\hat{\bf k}_3) Y_{\ell_2 m_2}(\hat{\bf k}_3) = \delta_{L,\ell_2}\delta_{M, m_2}$. Evaluating the second term in the same way, we finally obtain
\begin{eqnarray}
C_{\ell_1 m_1, \ell_2 m_2}^{\mu\mu}  &\approx&  
(-1)^{m_1} 16\pi \int \frac{k_1^2 dk_1}{2\pi^2} 
\int \frac{k_3^2 d k_3}{2 \pi^2} \nonumber \\
&& \left[ j_{\ell_1}^2 (k_3 x_{\rm ls}) + j_{\ell_2}^2(k_3 x_{\rm ls}) \right] \nonumber \\
&& f^2(k_1, k_1, k_3) P^2(k_1) h_{\ell_1 \ell_2 1}
\nonumber \\ 
&& 
\sum_M A_{1M}(k_1)
\left(
  \begin{array}{ccc}
  \ell_1 & \ell_2 & 1 \\
   -m_1 & m_2 & M 
  \end{array}
 \right) ~. \label{eq:mumu_ani_any_A}
\end{eqnarray}
Obviously, this recovers Eq.~\eqref{eq:mumu_ani_A} for $A(k) = const$.

\bibliography{paper}

\end{document}